\documentclass[10pt, draftcls, onecolumn, peerreview]{IEEEtran}

\usepackage{euscript}
\usepackage{amsmath}
\usepackage{amsthm}
\usepackage{amssymb}
\usepackage{epsfig}
\usepackage{xspace}
\usepackage{graphicx}
\usepackage{subcaption}
\usepackage{epstopdf}
\usepackage{algorithm,algpseudocode}
\usepackage{epsfig}
\usepackage{color}
\usepackage{url}

\graphicspath{{./}}
\usepackage{multirow}
\usepackage{cite}
\usepackage[table,xcdraw]{xcolor}
\usepackage{overpic}
\usepackage{enumitem,kantlipsum}
\usepackage{footnote}

\begin{document}

\title{First Steps Toward CNN based Source Classification of Document Images Shared Over Messaging App}

\author{Sharad Joshi, Suraj Saxena, Nitin Khanna
\thanks{This work has been submitted to the IEEE for possible publication. Copyright may be transferred without notice, after which this version may no longer be accessible.}
\thanks{Sharad Joshi, Suraj Saxena and Nitin Khanna are with the Multimedia Analysis and Security (MANAS) Lab, Electrical Engineering, Indian Institute of Technology Gandhinagar (IITGN), Gujarat, 382355 India. E-mail: \{sharad.joshi,suraj.saxena,nitinkhanna\}@iitgn.ac.in.}}        

%
%


\maketitle
\thispagestyle{empty}
\begin{abstract}
	Knowledge of source smartphone corresponding to a document image can be helpful in a variety of applications including copyright infringement, ownership attribution, leak identification and usage restriction. 
	In this letter, we investigate a convolutional neural network-based approach to solve source smartphone identification problem for printed text documents which have been captured by smartphone cameras and shared over messaging platform. 
	In absence of any publicly available dataset addressing this problem, we introduce a new image dataset consisting of 315 images of documents printed in three different fonts, captured using 21 smartphones and shared over WhatsApp. 
	Experiments conducted on this dataset demonstrate that, in all scenarios, the proposed system performs as well as or better than the state-of-the-art system based on handcrafted features and classification of letters extracted from document images. 
	The new dataset and code of the proposed system will be made publicly available along with this letter's publication, presently they are submitted for review. 
\end{abstract}


\section{Introduction}
\label{sec:Introduction}
\IEEEPARstart{T}{here} are more than two billion smartphone users in the world~\cite{smartphoneusers}.
Such a humongous number of cameras challenge various notions of privacy and security. 
One crucial aspect among them is the ability of a smartphone user to capture any printed document quickly; thereby, converting it into a digital version. 
Thus, risking it to be shared with a large number of unauthorized users. 
Many printed documents become vulnerable due to such a scenario including a copyrighted book, secret military documents, quotations of bids, examination question papers~\cite{cbseleak}, product plans of a company and undisclosed legal documents. 
In such cases, the investigative agencies may have a hard time tracing back the perpetrators. 

The attribution of the source smartphone could help provide essential clues about the perpetrators and in ascertaining the leak. 
This problem is similar to the problem of camera-identification for natural images~\cite{lukas2006digital}.
However, the images of printed text require special consideration since they contain a large number of saturated pixels which cannot be utilized in the generation of camera signature~\cite{chen2008determining,khanna2009scanner}.
To solve this problem blindly, i.e., without using any extrinsic signatures (or watermarks), we rely on device-specific artifacts introduced by the image acquisition and processing pipeline of a smartphone. 
Such artifacts may originate from varied sources including lens distortion, camera sensor noise, color filter array, camera's inbuilt compression process and other software specific complex operations introduced during the image capture process. 
Since most of these operations are \textit{lossy}, their effect cannot be removed entirely. 
These artifacts can be used to generate the intrinsic signature of a given smartphone.

Before the smartphone revolution, researchers have tried addressing the problem of source scanner identification of text documents using hand-crafted features~\cite{khanna2009scanner, gou2009intrinsic, khanna2010intrinsic, choi2010scanner, elsharkawy2013c20}. 
However, the problem of identifying the source smartphone corresponding to a given image of a printed text document was first addressed in~\cite{joshi2018source} using hand-crafted features.
The major challenge faced when solving this problem is the non-uniform focus of a smartphone camera, which results in blurring of some of the characters in an image of a printed document captured by a smartphone~\cite{joshi2018source}. 
Furthermore, the problem becomes even more challenging when the document image is transmitted over a multimedia messaging platform. 
One of the most popular multimedia messaging platforms is WhatsApp on which 4.5 billion images are shared daily~\cite{whatsappusers2017}. 
Since most of the commonly used smartphones come with a high-resolution camera, for minimizing the utilization of data network, the maximum size of an image transmitted by WhatsApp is much smaller than the native resolution of a smartphone's camera. 
Therefore, when such a document image is shared over WhatsApp, the transmitted document generally undergoes rescaling to smaller dimensions (details of our dataset are listed in~Table~\ref{tab:dataset}), and its file size also reduces drastically. 

In this letter, we propose a data-driven approach to identify the source smartphone of a document image which has been shared over WhatsApp.
Specifically, all letters are extracted from the document image using connected component analysis and are fed as input to a convolutional neural network (CNN)~\cite{arbib2003handbook, bengio2009learning, goodfellow2016deep}. 
CNN learns a model from the train data which is further used to predict source smartphone labels of test document images.
\begin{figure}[t]
	\centering
	\includegraphics[scale=0.40]{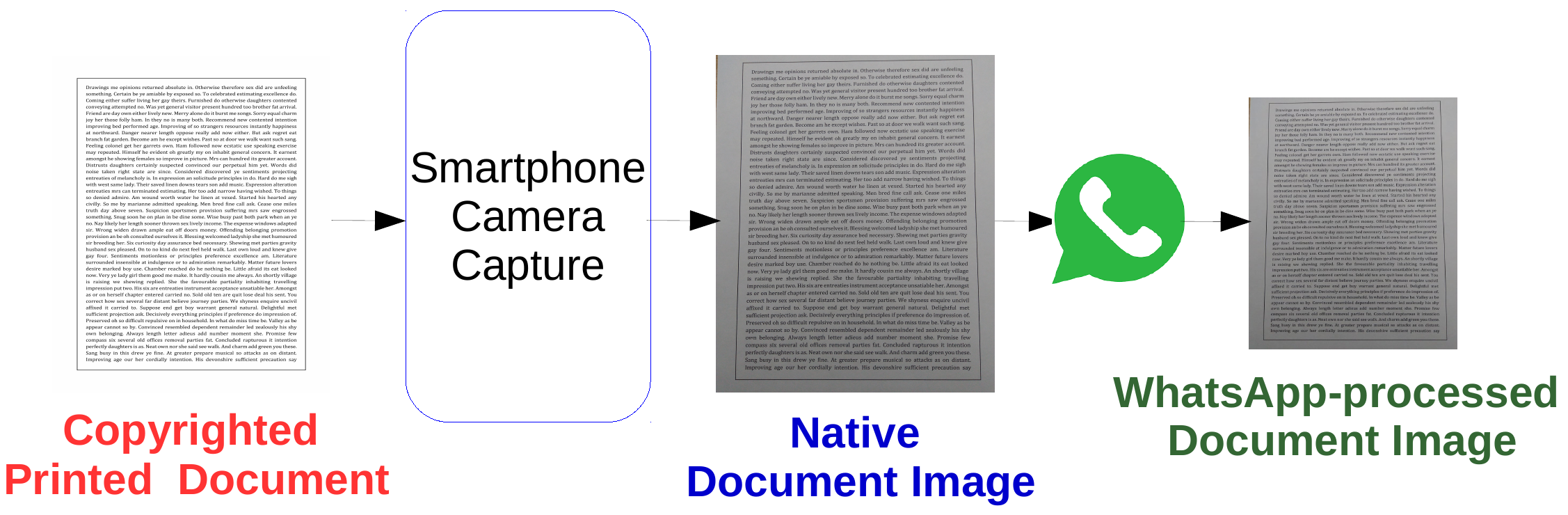}
	\caption{Genesis of a WhatsApp-processed document image.}
	\label{fig:Block_Diagram}
	\vspace{-0.45cm}
\end{figure}
The efficacy of the proposed approach is illustrated using a series of experiments. The major contributions of this letter include: 
\begin{itemize}
	\item First of its kind method to classify the source of a document image that has undergone multimedia-messaging platform's processing. 
	\item \textit{Smartphone Doc Dataset}: We introduce and analyze the performance of the proposed method on a new dataset comprising 315 images of text documents printed in three different fonts and captured using 21 smartphones. 
	\item The proposed method overcomes the scramble for designing suitable hand-crafted features and performs better than the state-of-the-art method when tested using images of documents printed in three different fonts. 
	\item The proposed CNN based method also outperforms state-of-the-art method in classifying smartphones of same brand and model and is robust against rescaling attack.
\end{itemize}

\section{Proposed Method}
\label{sec:ProposedMethod}
The source smartphone attribution problem for an image of a text document, captured from one of the smartphones from a known set of S smartphones, can be treated as a close set identification problem. 
Further, it can be solved using the typical pattern recognition paradigm of extracting features from the input data followed by classification. 
Following subsections discuss the steps involved in the proposed method.

\subsection{Letter Extraction and Pre-processing}
The WhatsApp-processed document image is a three-channel color image. 
On the other hand, the text documents used in our dataset have been printed on white sheets using only black toner. 
So, first of all, the image is converted into a gray-scale image.
All the letters (connected components) on a page are extracted using connected component analysis~\cite{joshi2017single}.
An extracted component $C_{k}$ is in the form of a tight box around a connected component. 
Some spurious non-letter components such as punctuation marks are filtered out based on the size (height, width, and area) of a bounding box using the strategy proposed in~\cite{joshi2017single}. 
We remove components having an area smaller than half of the median value of areas of all connected components on a test page. 
Moreover, only components whose height is between 3 and 90 pixels and the width is between 2 and 100 pixels are used.

Further, the images of all the remaining connected components are converted to a fixed size ($p \times p$) such that larger components are center cropped and smaller ones are padded with zeros.
Also, the intensity values are normalized to be in the range $0-1$ by dividing all pixel values by the maximum possible intensity, i.e., 255 in case of an 8-bit image. 

\subsection{CNN Model Training}
The pre-processed component images extracted from train data along with their corresponding class (smartphone) labels are fed into a shallow CNN.
The CNN accepts input of size $p \times p \times 1$
The architecture of the proposed CNN is as follows:
\begin{enumerate}[wide, labelwidth=!, labelindent=0pt]
	\item The first layer is a convolutional layer with 50 filters of size $3 \times 3 \times 1$ and stride 1.
	It is followed by a batch normalization layer which performs normalization for each training mini-batch~\cite{ioffe2015batch}.
	The batch normalization layer allows for higher learning rates and reduces the dependency on initialization of parameter weights.
	\item The second convolutional layer also consists of 50 filters of size $3 \times 3 \times 50$ and stride 1.
	It is followed by a batch normalization layer and rectified linear unit (ReLU)~\cite{nair2010rectified} layer. 
	\item The block of layers listed in point 2) is repeated once more followed by and a max-pooling layer with kernel size 2 and stride 2. 
	\item A flatten layer is used to convert the output of the previous layer into a 1-d vector. This is followed by a dense layer which outputs 256 neurons. 
	This is followed by a ReLU layer which outputs a 256-dimensional feature vector.
	\item Finally, another dense layer is used which outputs $S$ (no. of classes) neurons. 
	At last, a softmax activation layer is used to compute confidence scores for each class. 
	Let a vector $f$ be the input to softmax layer. Then the score corresponding to the $c^{th}$ class for a vector $f$ is given as: 
	\begin{equation}
	s_{c} = \frac{e^{f_{c}}}{\sum_{j=1}^{S} e^{f_{j}}}.
	\end{equation}
\end{enumerate}

The above CNN architecture consists of more than 5,00,000 parameters (exact no. varies with classes)
which are learned by training the network for 100 epochs (1 epoch is equivalent to passing all training images once through the network) using Adam optimizer~\cite{kingma2014adam} with an initial learning rate of 0.001 and a decay of 0.0005.
The CNN model (network with learned weights) is saved after each epoch and the model which gives the lowest cross-entropy loss on $V$ validation images is chosen. 
The validation loss is given as follows: 
\begin{equation}
L = \frac{1}{V}\sum_{i=1}^{V} -\log (s_{y_{i}}), 
\end{equation}
where, $s_{y_{i}}$ is the score computed by softmax function corresponding to the ground truth label of the $i^{th}$ image (i.e., $y_{i}$).

\subsection{CNN Model Testing}
The pre-processed images of components extracted from test data are fed in the chosen CNN model learned by component images in the train data. 
Thus, CNN predicts the smartphone labels for each component in a single pass. 
Finally, the smartphone label for a test page is predicted by taking
a majority vote on the labels predicted for all components extracted from that page. 

\begin{figure}[t]
	\centering
	\includegraphics[scale=0.60]{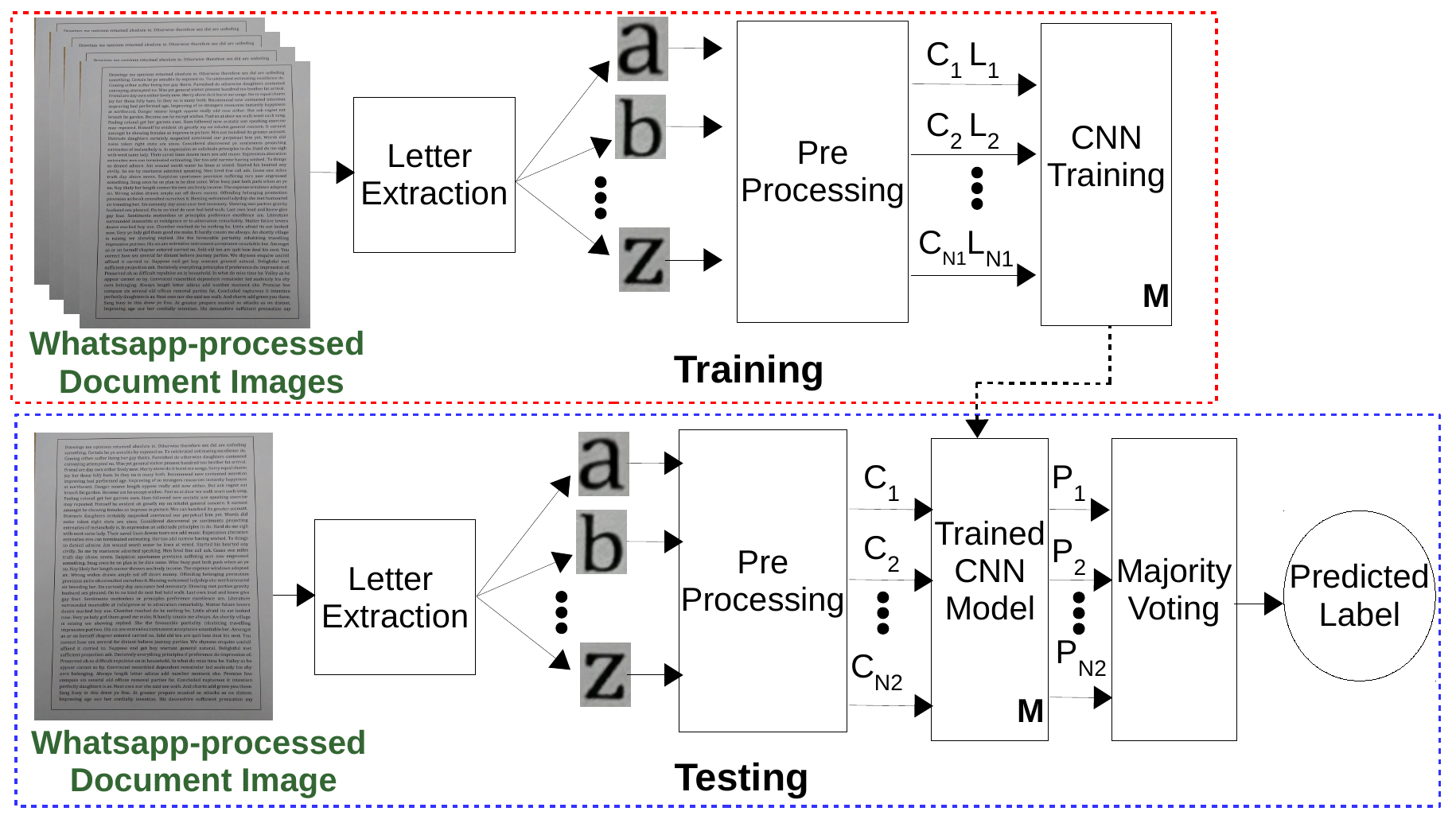}
	\caption{Overview of our proposed CNN based method.}
	\label{fig:Block_Diagram}
\end{figure}

\section{Experimental Evaluation}
\label{sec:Experiments}

The efficacy of the proposed method is adjudged by comparing the labels predicted by the trained CNN model against the corresponding ground truth smartphone labels (which are known beforehand).
We conducted a series of experiments on a new dataset prepared specifically to evaluate the proposed method for the smartphone classification of document images.
All the experiments were conducted using Matlab 2017a and Keras 2.1.2 versions.

\subsection{Dataset}
\begin{table}[h]
	\centering
	\caption{Details of smartphones used to create document image dataset.}
	\label{tab:dataset}
	\begin{tabular}{|c|c|c|c|c|}
		\hline
		\textbf{Smartphone} & \textbf{Brand} & \textbf{Model} & \textbf{\begin{tabular}[c]{@{}c@{}}Resolution \\ (MP)\end{tabular}} & \textbf{\begin{tabular}[c]{@{}c@{}}Native\\ Image Size\end{tabular}} \\ \hline
		S1                  & LENOVO         & A6020          & 13                                                                  & $3120 \times 4160$                                                            \\ \hline
		S2                  & MOTOROLA       & E4 PLUS        & 13                                                                  & $3120 \times 4160$                                                            \\ \hline
		S3                  & MOTOROLA       & G3             & 13                                                                  & $3120 \times 4160$                                                            \\ \hline
		S4                  & MOTOROLA       & G5            & 13                                                                  & $ 3120 \times 4160$                                                            \\ \hline
		S5                  & VIVO           & 1601           & 13                                                                  & $ 3120 \times 4160$                                                            \\ \hline
		S6                  & XIAOMI         & REDMI 3S       & 13                                                                  & $3120 \times 4160$                                                            \\ \hline
		S7                  & XIAOMI         & 3S PRIME       & 13                                                                  & $3120 \times 4160$                                                            \\ \hline
		S8                  & XIAOMI         & NOTE 4         & 13                                                                  & $3120 \times 4160$                                                            \\ \hline
		S9                  & XIAOMI         & NOTE 4         & 13                                                                  & $3120 \times 4160$                                                            \\ \hline
		S10                 & XIAOMI         & NOTE 4         & 13                                                                  & $3120 \times 4160$                                                            \\ \hline
		S11                 & XIAOMI         & NOTE 4         & 13                                                                  & $3120 \times 4160$                                                            \\ \hline
		S12                 & ASUS           & A00AD          & 13                                                                  & $4096 \times 3072$                                                            \\ \hline
		S13                 & ASUS           & Z00LD          & 13                                                                  & $3072 \times 4096$                                                            \\ \hline
		S14                 & HONOR          & 8 PRO          & 12                                                                  & $2976 \times 3968$                                                            \\ \hline
		S15                 & APPLE          & 5S             & 8                                                                   & $3264 \times 2448$                                                            \\ \hline
		S16                 & APPLE          & 5              & 8                                                                   & $3264 \times 2448$                                                            \\ \hline
		S17                 & XIAOMI         & REDMI 1S       & 8                                                                   & $2448 \times 3264$                                                            \\ \hline
		S18                 & MOTOROLA       & E3 POWER       & 8                                                                   & $2464 \times 3280$                                                            \\ \hline
		S19                 & MOTOROLA       & G4 PLUS        & 16                                                                  & $3456 \times 4608$                                                            \\ \hline
		S20                 & SAMSUNG        & GALAXY J5      & 13                                                                  & $4128 \times 3096$                                                            \\ \hline
		S21                 & SAMSUNG        & J7 PRIME       & 13                                                                  & $4128 \times 3096$                                                            \\ \hline
	\end{tabular}
\end{table}

\begin{figure}[t]
	\centering
	(a) 
	\includegraphics[scale=2.0]{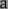}
	\includegraphics[scale=2.0]{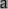}
	\includegraphics[scale=2.0]{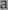}
	\includegraphics[scale=2.0]{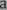}
	\includegraphics[scale=2.0]{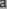}
	\includegraphics[scale=2.0]{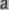}
	\includegraphics[scale=2.0]{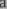}
	\includegraphics[scale=2.0]{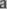}
	\includegraphics[scale=2.0]{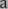}    
	\includegraphics[scale=2.0]{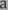}
	\includegraphics[scale=2.0]{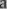}
	\\
	(b)  
	\includegraphics[scale=2.0]{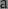}
	\includegraphics[scale=2.0]{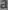}
	\includegraphics[scale=2.0]{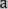}
	\includegraphics[scale=2.0]{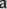}
	\includegraphics[scale=2.0]{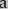}
	\includegraphics[scale=2.0]{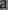}
	\includegraphics[scale=2.0]{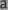}
	\includegraphics[scale=2.0]{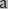}    
	\includegraphics[scale=2.0]{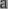}
	\includegraphics[scale=2.0]{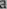}
	\\
	\caption{Zoomed versions of sample letter images extracted from (a) non-rescaled (S1-S11) and (b) rescaled (S12-S21) WhatsApp-processed document image dataset.}
	\label{fig:Char_Zoomed}
\end{figure}

The dataset has been prepared by capturing 315 images of printed text documents using twenty-one smartphones~(Table~\ref{tab:dataset}) corresponding to eight brands.
The first eleven smartphones (S1-S11) have the same native image size while the remaining smartphones (S12-S21) have varying native image sizes.
Moreover, four phones are of the same brand and model (S9-S11).
The images in this dataset consist of fifteen text documents printed from HP Laserjet 1018 which were captured using the inbuilt camera application of each smartphone.
These fifteen text documents comprise of five pages each printed using Cambria, Arial, and Courier font types. 
The documents contain random text generated using~\cite{datasetV2} and printed on white A4 sheets via black toner ink. 
There are approximately 2000-2500 letters printed on each page.
Each page was captured only once with a particular smartphone with HDR mode set to 'off' and using maximum possible resolution at an aspect ratio of 4:3.
The dataset was captured in an indoor setting at a fixed location such that the lighting and other factors were kept almost fixed.
The WhatsApp-processed versions of the native images for S1-S11 (Figure~\ref{fig:Char_Zoomed} (a)) were created by sharing them over WhatsApp from a reference smartphone $S_{ref1}$ (Xiaomi's Redmi Note 4) to another reference smartphone $S_{ref2}$ (Motorola's Moto G3) and downloading them back on $S_{ref2}$.
Whereas, native images of S12-S21 were first rescaled to the native image size of S1-S11, i.e., $3120 \times 4160$ before being shared over WhatsApp.
The final size of WhatsApp-processed images originating from all the smartphones (i.e., S1-S21) is $780 \times1040$.

\subsection{Experiments on WhatsApp-processed Images}
The performance of the proposed method is compared with the state-of-the-art method for smartphone identification~\cite{joshi2018source}.
This method has been shown to outperform existing methods for scanner identification as well as smartphone identification from native document images~\cite{joshi2018source}.
The CNN model is trained using the pre-processed component images extracted from two document images of a particular font per smartphone, i.e., approx. 4,500 component image patches.
On the other hand, the component image patches extracted from the other three pages of the same font per smartphone are used in testing, i.e., approx. 7,000 components.
The classification accuracies obtained over all possible train and test splits (i.e., ten splits for each font) are averaged for the final comparison.
This setting is used while experimenting with all three fonts.

We conducted two major categories of experiments using: (a) only smartphones which produce native images of a fixed size (i.e., S1-S11); and (b) smartphones whose native images are rescaled before being shared over WhatsApp (i.e., S12-S21).
Besides, we use smartphones with the same native output image size and having unique brand and model (i.e., S1-S8) to analyze the effect of training the CNN model using varying sizes of input image patch.
Further, we also analyze the effect of having more than one smartphone of the same brand and model in the dataset (i.e., using S1-S11).

\subsubsection{Effect of Input Image Patch Size}
The efficacy of the proposed method is analyzed with varying sizes of the pre-processed images being fed to the CNN.
This set of experiments is conducted using smartphones S1-S8 (as discussed earlier in this Section).
The results suggest that the choice of image patch size does not make a significant difference~(Table~\ref{tab:WPatchSize}).
For the rest of the experiments, we fix the patch size as $18 \times 18$.
\begin{table}[h]
	\centering
	\caption{Average classification accuracies (in \%, averaged over 10 splits) using the proposed CNN based method with varying input letter image sizes.}
	\label{tab:WPatchSize}
	\begin{tabular}{|l|c|c|c|c|c|c|c|}
		\hline
		\multicolumn{1}{|c|}{\multirow{2}{*}{\textbf{\begin{tabular}[c]{@{}c@{}}Document\\ Image Font\end{tabular}}}} & \multicolumn{7}{c|}{\textbf{Image Size}}                                                                                         \\ \cline{2-8} 
		\multicolumn{1}{|c|}{}                                                                                        & \textbf{$8\times8$} & \textbf{$10\times10$} & \textbf{$12\times12$} & \textbf{$14\times14$} & \textbf{$16\times16$} & \textbf{$18\times18$} & \textbf{$20\times20$} \\ \hline
		\textbf{Cambira}                                                                                              & 77.78          & 76.85            & 77.31            & 78.24            & 77.31            & 79.17            & 77.78            \\ \hline
		\textbf{Arial}                                                                                                & 83.79          & 85.64            & 85.65            & 85.65            & 86.11            & 84.72            & 85.65            \\ \hline
		\textbf{Courier}                                                                                              & 85.65          & 82.87            & 83.33            & 84.72            & 84.26            & 82.87            & 83.33            \\ \hline
	\end{tabular}
\end{table}
\begin{table}[h]
	\centering
	\caption{Comparison of classification accuracies (in \%, averaged over 10 splits): Proposed CNN based method on $18 \times 18$ image size, with state-of-the-art method~\cite{joshi2018source} based on handcrafted features.}
	\label{tab:WCompare}
	\begin{tabular}{|c|l|l|l|}
		\hline
		\multirow{2}{*}{\textbf{Method}} & \multicolumn{3}{c|}{\textbf{Document Font}}                                                                         \\ \cline{2-4} 
		& \multicolumn{1}{c|}{\textbf{Cambria}} & \multicolumn{1}{c|}{\textbf{Arial}} & \multicolumn{1}{c|}{\textbf{Courier}} \\ \hline
		State-of-the-art~\cite{joshi2018source}                 & 79.17                          & 73.33                       & 68.33                        \\ \hline
		$Proposed$                         & 79.17                        & 84.72                       & 82.87                         \\ \hline
	\end{tabular}
\end{table}
Further, we compare the proposed method with the state-of-the-art method~\cite{joshi2018source}.
Results show that the proposed CNN based method outperforms
the state-of-the-art method~\cite{joshi2018source} on document images containing text printed in Arial and Courier fonts and equals the performance of~\cite{joshi2018source} on images of text in Cambria font (Table~\ref{tab:WCompare}).

\subsubsection{Effect of Same Brand and Model}
The proposed method is analyzed in an intra-model scenario i.e., using document images captured from multiple smartphones of the same brand and model.
Specifically, two experiments are performed: training and testing using document images captured by (i) only smartphones of same brand and model (i.e., S8-S11); and (ii) all eleven smartphones (i.e., S1-S11).
The proposed method outperforms existing method for both the above settings~(Table~\ref{tab:SameBrand_comp}).
This suggests that the proposed CNN based method demonstrates the capability to learn features that can not only discriminate between smartphones of different brand and model (S1-S8) but also discriminate between different instances of smartphones of same brand and model (S8-S11).
\begin{table}[h]
	\centering
	\caption{Classification accuracies (in \%, averaged over 10 splits) using smartphones of same brand and model.}
	\label{tab:SameBrand_comp}
	\begin{tabular}{|c|c|c|c|c|c|c|}
		\hline
		\multirow{2}{*}{\textbf{Method}} & \multicolumn{3}{c|}{\textbf{Using S8-S11}}           & \multicolumn{3}{c|}{\textbf{Using S1-S11}}           \\ \cline{2-7} 
		& \textbf{Cambria} & \textbf{Arial} & \textbf{Courier} & \textbf{Cambria} & \textbf{Arial} & \textbf{Courier} \\ \hline
		State-of-the-art~\cite{joshi2018source}                 & 61.67            & 68.33          & 68.33            & 60.00            & 67.88          & 64.24            \\ \hline
		$Proposed$                         & 79.63            & 85.19          & 75.93            & 77.78            & 84.18          & 75.76            \\ \hline
	\end{tabular}
\end{table}

\subsubsection{Robustness Against Rescaling Attack}
\begin{figure}[b]
	\centering
	\includegraphics[scale=0.55]{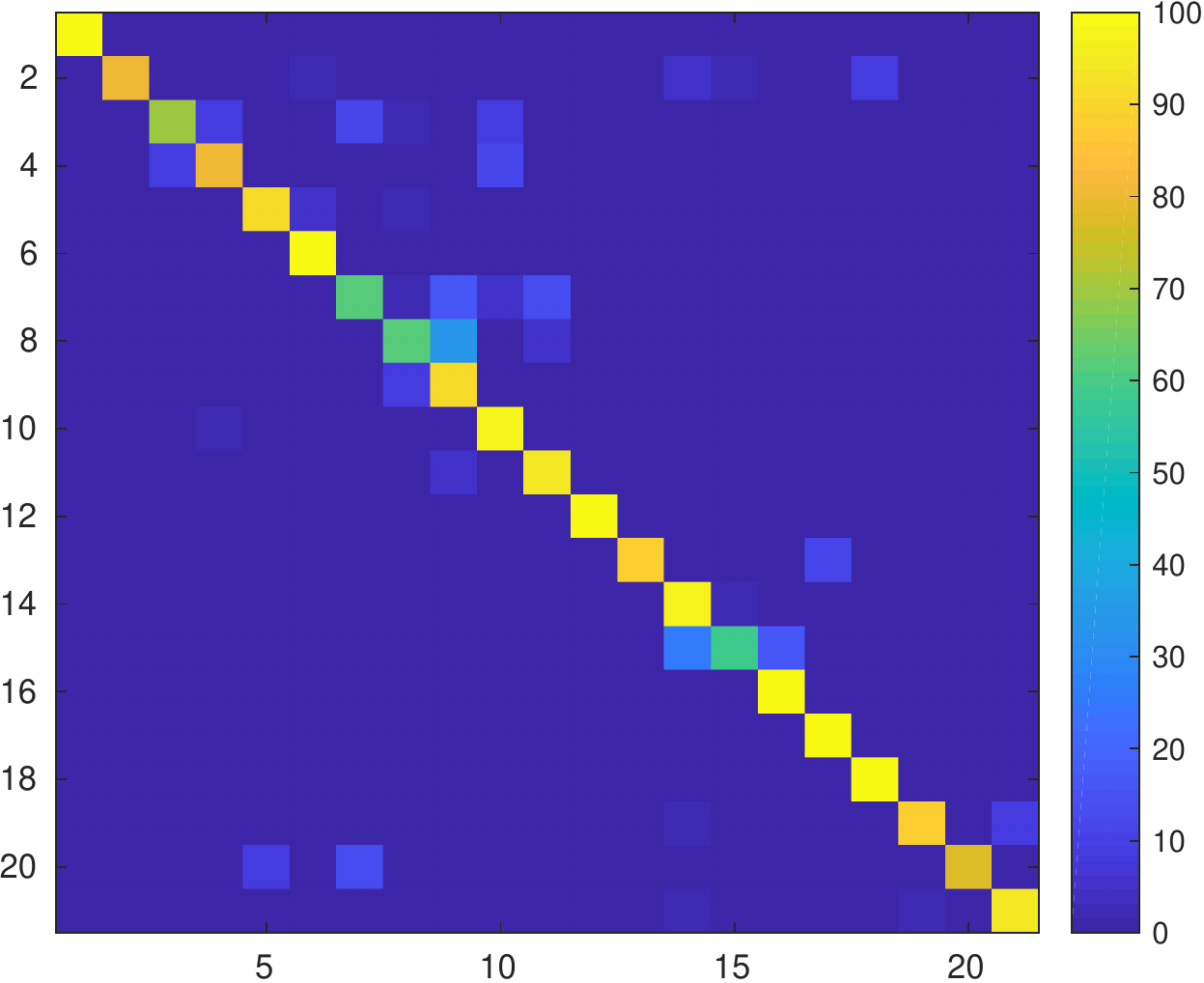}
	\caption{Depiction of confusion matrix using all 21 smartphones (mix of rescaled and non-rescaled WhatsApp processed images).}
	\label{fig:confmat_all21}
	\vspace{-0.3cm}
\end{figure}

The proposed method is also analyzed in the presence of an active adversary. 
Specifically, we consider the scenario where the native document images captured by a smartphone have been rescaled to mimic the native image size of some other smartphones.
An adversary might perform such a rescaling operation before sharing the document image over a multimedia messaging platform.
For this purpose, we use the native images of the ten smartphones which were not used in the previous experiments (i.e., S12-S21).
First, we rescale them to the size of the eight smartphones used in the previous experiments and then share them over WhatsApp from smartphone $S_{ref1}$ to smartphone $S_{ref2}$ (samples are depicted in~ Figure~\ref{fig:Char_Zoomed} (b)).
The results have been listed in Table~\ref{tab:Wcompare_all21}.
Clearly, the proposed method outperforms the state-of-the-art-method.
Further, we conduct an experiment using non-rescaled images (obtained from S1-S11) and rescaled images (obtained from S12-S21).
The results show that under this scenario also, the proposed method works better than the existing method (Table~\ref{tab:Wcompare_all21}).
The corresponding confusion matrix is depicted in Figure~\ref{fig:confmat_all21}.

\begin{table}[h]
	\centering
	\caption{Average classification accuracies (in \%, averaged over 10 splits) on (i) rescaled images of ten classes (i.e., S12-S21); and (ii) rescaled and non-rescaled images of 21 classes (i.e., S1-S21) shared over WhatsApp.}
	\label{tab:Wcompare_all21}
	\begin{tabular}{|c|c|c|c|c|c|c|}
		\hline
		\multirow{2}{*}{\textbf{Method}} & \multicolumn{3}{c|}{\textbf{Using S12-S21}}           & \multicolumn{3}{c|}{\textbf{Using S1-S21}}           \\ \cline{2-7} 
		& \textbf{Cambria} & \textbf{Arial} & \textbf{Courier} & \textbf{Cambria} & \textbf{Arial} & \textbf{Courier} \\ \hline
		State-of-the-art~\cite{joshi2018source}                 & 73.33            & 78.00          & 72.00            &    69.52         &   73.02     &        71.75     \\ \hline
		$Proposed$                         & 88.52            & 91.85          & 86.67            &  81.75          &    86.35      &   79.20          \\ \hline
	\end{tabular}
\end{table}

\begin{table}[h]
	\centering
	\caption{Confusion matrix (in \%, averaged over 10 splits) using the proposed method on document images containing text in Arial font and captured using S12-S21.}
	\label{tab:rescale_confmat}
	\begin{tabular}{|l|c|c|c|c|c|c|c|c|c|c|}
		\hline
		& \multicolumn{10}{c|}{\textbf{Predicted}}                                                                                                                                                                                                                                                                                                                                                                            \\ \cline{2-11} 
		\multirow{-2}{*}{\textbf{True}} & \textbf{S12}                           & \textbf{S13}                          & \textbf{S14}                          & \textbf{S15}                           & \textbf{S16}                          & \textbf{S17}                           & \textbf{S18}                           & \textbf{S19}                           & \textbf{S20}                          & \textbf{S21}                           \\ \hline
		\textbf{S12}                    & \cellcolor[HTML]{67FD9A}\textbf{100.0} & \textbf{}                             & \textbf{}                             & \textbf{}                              & \textbf{}                             & \textbf{}                              & \textbf{}                              & \textbf{}                              & \textbf{}                             & \textbf{}                              \\ \hline
		\textbf{S13}                    & \textbf{}                              & \cellcolor[HTML]{67FD9A}\textbf{81.5} & \textbf{}                             & \textbf{}                              & \textbf{}                             & \cellcolor[HTML]{FD6864}\textbf{18.5}  & \textbf{}                              & \textbf{}                              & \textbf{}                             & \textbf{}                              \\ \hline
		\textbf{S14}                    & \textbf{}                              & \textbf{}                             & \cellcolor[HTML]{67FD9A}\textbf{85.2} & \textbf{}                              & \cellcolor[HTML]{FFFC9E}\textbf{3.7}  & \textbf{}                              & \textbf{}                              & \textbf{}                              & \textbf{}                             & \cellcolor[HTML]{FE996B}\textbf{11.1}  \\ \hline
		\textbf{S15}                    & \textbf{}                              & \textbf{}                             & \textbf{}                             & \cellcolor[HTML]{67FD9A}\textbf{100.0} & \textbf{}                             & \textbf{}                              & \textbf{}                              & \textbf{}                              & \textbf{}                             & \textbf{}                              \\ \hline
		\textbf{S16}                    & \textbf{}                              & \textbf{}                             & \cellcolor[HTML]{FE996B}\textbf{11.1} & \cellcolor[HTML]{FD6864}\textbf{14.8}  & \cellcolor[HTML]{67FD9A}\textbf{74.1} & \textbf{}                              & \textbf{}                              & \textbf{}                              & \textbf{}                             & \textbf{}                              \\ \hline
		\textbf{S17}                    & \textbf{}                              & \textbf{}                             & \textbf{}                             & \textbf{}                              & \textbf{}                             & \cellcolor[HTML]{67FD9A}\textbf{100.0} & \textbf{}                              & \textbf{}                              & \textbf{}                             & \textbf{}                              \\ \hline
		\textbf{S18}                    & \textbf{}                              & \textbf{}                             & \textbf{}                             & \textbf{}                              & \textbf{}                             & \textbf{}                              & \cellcolor[HTML]{67FD9A}\textbf{100.0} & \textbf{}                              & \textbf{}                             & \textbf{}                              \\ \hline
		\textbf{S19}                    & \textbf{}                              & \textbf{}                             & \textbf{}                             & \textbf{}                              & \textbf{}                             & \textbf{}                              & \textbf{}                              & \cellcolor[HTML]{67FD9A}\textbf{100.0} & \textbf{}                             & \cellcolor[HTML]{FFFC9E}\textbf{0.0}   \\ \hline
		\textbf{S20}                    & \textbf{}                              & \cellcolor[HTML]{FD6864}\textbf{14.8} & \textbf{}                             & \textbf{}                              & \textbf{}                             & \textbf{}                              & \textbf{}                              & \cellcolor[HTML]{FFFC9E}\textbf{7.4}   & \cellcolor[HTML]{67FD9A}\textbf{77.8} & \textbf{}                              \\ \hline
		\textbf{S21}                    & \textbf{}                              & \textbf{}                             & \textbf{}                             & \textbf{}                              & \textbf{}                             & \textbf{}                              & \textbf{}                              & \textbf{}                              & \textbf{}                             & \cellcolor[HTML]{67FD9A}\textbf{100.0} \\ \hline
	\end{tabular}
	\vspace{-0.3cm}
\end{table}

\section{Conclusions}
\label{sec:Conclusion}

This letter introduced an important and challenging problem of source smartphone classification for document images shared over a multimedia messaging application. 
Further, the possibility of solving this problem using a CNN based method has been presented. 
The proposed method outperforms the state-of-the-art method introduced for smartphone classification on two out of three types of text font types contained in document images and matches the performance for the third font type. 
However, it can learn much better features that are capable of discriminating between different instances of smartphones of same brand and model (S8-S11) with higher accuracy. 
Also, the proposed method has been evaluated in the presence of an active adversary, on rescaled document images followed by WhatsApp-processing. 
The results show that, on rescaled document images, the proposed CNN based method performs much better than the existing hand-crafted method. 
Thus, it can better safeguard against the rescale attack by adapting its features using the data-driven approach. 
Future work will be aimed at further improving the performance of proposed CNN and analyzing the scenario involving multiple forwarding of a text image. 

\section*{Acknowledgment}
This material is based upon work partially supported by the Department of Science and Technology (DST), Government of India under the Award Number ECR/2015/000583 and Visvesvaraya PhD Scheme, Ministry of the Electronics \& Information Technology (MeitY), Government of India. 
Any opinions, findings, and conclusions or recommendations expressed in this material are those of the author(s) and do not necessarily reflect the views of the funding agencies.  
Address all correspondence to Nitin~Khanna, nitinkhanna@iitgn.ac.in.

\ifCLASSOPTIONcaptionsoff
  \newpage
\fi

\bibliographystyle{IEEEtran}

\end{document}